\documentclass[12pt]{article} 
\usepackage{amsfonts, amssymb, amsmath}

\newtheorem{thm}{Theorem} 

 \def\be{\begin{eqnarray}}
\def\ee{\end{eqnarray}}
\def\bee{\begin{eqnarray*}}
\def\eee{\end{eqnarray*}}

 \def\pmx{\begin{pmatrix}}
 \def\emx{\end{pmatrix}}
 \def\bsq{\begin{subequations}}
\def\esq{\end{subequations}}

\def\raw{\rightarrow}

    \def\nn{\nonumber}
\newcommand{\norm}[1]{ \| #1  \|}
 \def\half{{\textstyle \frac{1}{2}}}
 \def\dtsig{{\mathbf \cdot \sigma}}
\def\bw{{\bf w}}

        \def\tr{\hbox{Tr} \, }
     
     \def\av{{\rm Av}}
     \def\ren{{\rm Ren}}
     \def\ot{\otimes}
     \def\wtd{\widetilde}
     
     \def\rev{{\rm Rev}}
    
\def\bs{\bigskip}
    
  \title{{\LARGE Bounds on the Concavity of Quantum Entropy}}
  
  \author{Isaac Kim  \\ Perimeter Institute, Waterloo, Ontario, Canada N2L 2Y5 \\ 
  {\small ikim@perimeterinstitute.ca} \\~~ \\
  \and Mary Beth Ruskai \\
  Institute for Quantum Computing,  University of Waterloo \\ Waterloo, Ontario, Canada  N2L 3G1\\  {\small ruskai@member.ams.org}}  
  
  \date{  \today \\ ~~ \\{\em  Dedicated to the memory of Oscar E. Lanford III}}
  
\begin{document}

\maketitle

\begin{abstract}
We give new upper and lower bounds on the concavity of quantum entropy.
Comparisons are given with other results in the literature.
\end{abstract}

\section{Introduction}

It is well-known that the von Neumann quantum entropy  $S(\rho) \equiv - \tr \rho \log \rho $ is 
concave, where the argument $\rho$ is a density matrix $\rho$, i.e., $\rho \geq 0, ~ \tr \rho = 1$.  
Recently Kim \cite{kim} proved the following lower bound on the concavity
\be     \label{lowbd0}
  S(   \rho_\av) - x S(\rho_1) - (1-x) S(\rho_2) & \geq &  \frac{x(1-x)}{(1-2x)^2 }
      \max \begin{cases}   H(\rho_\av,  \rho_\rev ) \\  H( \rho_\rev, \rho_\av)  \end{cases}
     \\     \label{lowbd1}
    & \geq & \half x(1-x)  \norm{\rho_1- \rho_2}_1^2
\ee
where  $\rho_\av \equiv x \rho_1 + (1-x) \rho_2, ~ \rho_\rev   \equiv x \rho_2 + (1-x) \rho_1 $
and   $H(\rho,\gamma) \equiv \tr \rho \big(\log \rho - \log \gamma )$ denotes the relative entropy.
In Section~\ref{sect:lower} present a simple proof of \eqref{lowbd1} and compare it to other possible lower bounds, particularly those resulting from recent work of Carlen and Lieb \cite{CL}.
Our results demonstrate that their methods do  not always give stronger results than those obtained using Pinsker's bound.
We also consider upper bounds in Section~\ref{sect:up}.  Finally, in section~\ref{sect:renyi} we discuss the possibility of improved bounds using variants of the Renyi relative entropy.

\section{Lower bounds}   \label{sect:lower}
We begin with a simple proof of \eqref{lowbd1}
\be \lefteqn{   S(\rho_\av) - x S(\rho_1) - (1-x) S(\rho_2)   }    \qquad \qquad \quad  \qquad \nn  \\
   & =   & x \tr \rho_1 \log \rho_1 + (1-x) \tr \rho_2 \log \rho_2 - \tr [x \rho_1 + (1-x) \rho_2] \log \rho_\av   \nn  \\
   & =   & xH( \rho_1 , \rho_\av)+ (1-x) H(\rho_2 , \rho_\av)  \\
   & \geq & \half x \norm{ \rho_1 - \rho_\av }_1^2  +  \half (1-x) \norm{ \rho_2 - \rho_\av }_1^2  \nn \\
   & = &  x (1-x) \half \norm{\rho_1 - \rho_2}_1^2   
\ee
where we used Pinsker's bound  \cite{P} \cite[Theorem 1.15]{OP}
\be
   H(\rho,\gamma) \geq \half \norm{\rho - \gamma}_1^2
\ee

In view of the fact that Carlen-Lieb \cite{CL} presented an inequality for subadditivity that can be stronger than Pinsker's  bound in some situations, it is interesting to see if one can use their results to   improve the bound \eqref{lowbd1}.
We begin with the well-known observation that if
\be   \label{block}
    P_{AB} = \pmx x \rho_1 & 0 \\ 0 & (1-x) \rho_2 \emx
\ee
so that $P_A = \rho_\av$ and $P_B = \pmx x & 0 \\ 0 & 1-x \emx $, then
\be   \label{subadd}
      H(P_{AB}, P_A \ot P_B) & = &   S(P_A) +S(P_B) - S(P_{AB}) \nn \\
            & = &  S( \rho_\av) - x S(\rho_1) - (1-x) S(\rho_2) 
\ee
It then follows from \cite[Lemma~2.1]{CL} that
\be     \allowdisplaybreaks
 \lefteqn{  S( \rho_\av) - x S(\rho_1) - (1-x) S(\rho_2) }  \qquad \qquad \nn \\
 & \geq & - 2 \log \Big[ 1 - \half \tr \big( \sqrt{P_{AB} }- \sqrt{P_A \ot P_B} \big)^2 \Big] \\ \nn 
 & = & - 2 \log \Big[ 1 - \half \tr 
       \big(P_{AB} + P_A \ot P_B -  2 \sqrt{ P_{AB}} \sqrt{P_A\ot P_B}  \big) \Big]  \nn \\
     & = & - 2 \log \tr  \sqrt{ P_{AB} }    \sqrt{P_A\ot P_B }\big)  \label{lowbd2}  \\
     &  = & - 2 \log \tr \pmx   x \sqrt{\rho_1} \sqrt{\rho_\av} & 0 \\  0 & (1-x) \sqrt{\rho_2}\sqrt{\rho_\av }\emx  \nn
      \allowdisplaybreaks  \\
     & = & -  2\log \tr  \big[  x \sqrt{\rho_1}  +  (1-x)  \sqrt{\rho_2} \big]  \sqrt{\rho_\av}   \label{lowbd3} \\
     & \geq & - 2 \log \tr    \sqrt{\rho_\av}    \sqrt{\rho_\av}  = 0   \nn 
\ee
where the last inequality uses the fact that $f(u) =\sqrt{u} $ is a concave operator function.
Note that the bound  \eqref{lowbd2} could have been obtained directly from the montonicity of
the Renyi relative entropy  
\be  \label{def:Renyi}
H_a^\ren(\rho,\gamma) \equiv \frac{1}{a-1} \log \tr \rho^a \gamma^{1-a} 
\ee
in the case 
\be
   H(\rho,\gamma) = H_1^\ren(\rho,\gamma) \geq H_{1/2}^\ren(\rho,\gamma)
\ee
If one applies Pinsker's inequality to \eqref{subadd}, one would obtain
\be
    S( \rho_\av) - x S(\rho_1) - (1-x) S(\rho_2)  =  H(P_{AB}, P_A \ot P_B)  \geq 2 x^2(1-x)^2
\ee
which appears to be weaker than \eqref{lowbd2}.   

This raises the question of which of the lower bounds on concavity is stronger.   Numerical
tests show that in some cases \eqref{lowbd1} is stronger and in other cases \eqref{lowbd2}  is stronger.
(See the Appendix for specific examples.)
Thus the Carlen-Lieb bound \cite{CL} on subadditivity does not give a stronger bound on the concavity of
 entropy than Pinsker's inequality.

 \section{Upper bounds}  \label{sect:up}
 
It is also a well-known consequence  \cite{DM,LanR}
of the operator monotonicity of the function $f(u) = \log u$ that the concavity satisfies the upper bound
\be   \label{upbd}
    S(   \rho_\av) - x S(\rho_1) - (1-x) S(\rho_2) \leq - x \log x - (1-x) \log (1-x) \equiv h(x)
\ee
where $h(x)$ denotes the binary entropy.  Using the Bures metric\footnote{ $\tr \big( \sqrt{\rho} \, \gamma \,  \sqrt{\rho} \big)^{1/2} $ 
 is known as the fidelity in quantum information theory.} \cite{Bur,U} 
\be  
  D^2_{\rm Bures}(\rho ,\gamma )   \equiv  2 \Big[1 -  \tr  \big( \sqrt{\rho}\, \gamma \,  \sqrt{\rho} \big)^{1/2} \Big] 
    \leq \norm{\rho - \gamma}_1
\ee
with the inequality due to Fuchs and van de Graph \cite{FG}, the inequality 
\be     \label{RFZ}
     S(   \rho_\av) - x S(\rho_1) - (1-x) S(\rho_2) \leq h(x) D_{\rm Bures}^2(\rho_1,\rho_2)     \leq  h(x) \norm{\rho_1 - \rho_2}_1
\ee
 was proved in \cite{RFZ}.  However, because one can have $ \norm{\rho_1 - \rho_2}_1 > 1$,
\eqref{RFZ} is not always stronger than \eqref{upbd}.

Recently Audenaert  \cite[Eq.~66]{A}  obtained the stronger upper bound
\be
      \label{upA}
     S(   \rho_\av) - x S(\rho_1) - (1-x) S(\rho_2) \leq h(x) \half \norm{\rho_1 - \rho_2 }_1   
\ee
which is also stronger than \eqref{upbd}. 

A different upper bound is given in Theorem~\ref{thm:milan} below.

 \section{Renyi bounds}  \label{sect:renyi}
We can summarize the basic bounds as 
 \begin{thm}   \label{thm1}
 The quantum entropy satisfies the upper and lower bounds
   \be
   h(x) \half \norm{\rho_1 - \rho_2 }_1    & \geq & 
       S(   \rho_\av) - x S(\rho_1) - (1-x) S(\rho_2) \nn \\   & \geq &
         \half x(1-x)  \norm{\rho_1- \rho_2}_1^2   \ee
   \end{thm}
It is natural to ask if one can use the monotonicity of either the Renyi relative entropy \eqref{def:Renyi}
or the  sandwiched Renyi entropy 
\be    \label{def:sand}
   \wtd{H}^\ren_a(\rho , \gamma) \equiv   \tr \frac{1}{a-1} \log \tr \Big(\gamma^{\frac{1-a}{2a}} \rho  \gamma^{\frac{1-a}{2a}} \Big)^a
\ee
introduced independently in \cite{MDSFT} and \cite{WWY}
 to improve these bounds.   It is well-known that the Renyi relative entropy \eqref {def:Renyi}
   is monotone in $a$ (see, e.g., \cite{M,MDSFT} ) and Beigi \cite{B}  (see also \cite{MDSFT}) recent proved
   that this monotonicity also holds for the
sandwiched Renyi entropy \eqref{def:sand}  when $a \geq \half $.
 Because  $  \wtd{H}^\ren_a(\rho , \gamma)  \leq H^\ren_a(\rho,\gamma) ~~ \forall ~ a $, it is always more advantageous to use 
   $  \wtd{H}^\ren_a $ for upper bounds and $H^\ren_a  $ for lower bounds.   Since
   \be
         \lim_{a \raw 1} \wtd{H}^\ren_a(\rho , \gamma) =   H (\rho,\gamma)  =  \lim_{b \raw 1} H^\ren_b(\rho,\gamma) 
   \ee 
   it follows that
   \begin{thm}  
   For any fixed, $x \in (0,1)$ and $ \rho_1 \neq  \rho_2 $ one can find  $a_c > 1$ and $b_c \in  [\half,1) $ such that
 \bee
    h(x) \half \norm{\rho_1 - \rho_2 }_1 & \geq &  x   \wtd{H}^\ren_a(\rho_1 , \rho_\av)   + (1-x)  \wtd{H}^\ren_a(\rho_2 , \rho_\av) 
   \quad \forall ~ a \geq a_c \nn \\
  & \geq & 
       S(   \rho_\av) - x S(\rho_1) - (1-x) S(\rho_2) \\    \nn
       & \geq &
       x H^\ren_b (\rho_1 , \rho_\av)  + (1-x)  H^\ren_b(\rho_2 , \rho_\av)    \quad \forall ~ b \in [b_c,1) \quad  \\ \nn
      & \geq &   \half x(1-x)  \norm{\rho_1- \rho_2}_1^2
        \eee
   \end{thm}
   Except for very special situations, e.g., both $\rho_1, \rho_2$ multiples of orthogonal projections, one would expect that 
   one can find $a_c, b_c$ such that all of the above bounds are strict.
   However, it seems unlikely that there exist $a$ and/or $b$ which improve the basic bounds in Theorem~1 
   for arbitrary $x, \rho_1, \rho_2 $.   Whether or not one can obtain improved bounds if some of these parameters
   are fixed seems to be an open question in general.  
   
   In the special case $x = \half$, observe that
     \be
      \half \Big[H^\ren_2 (\rho_1 , \rho_\av)  +   H^\ren_2(\rho_2 , \rho_\av) \Big]  
         & = & \half  \Big[ \log   \tr \rho_1^2 \rho_\av^{-1} + \log   \tr \rho_2^2 \rho_\av^{-1} \Big] \nn \\
         & \leq & \half  \big[ \log ( 2 \,  \tr \rho_1) + \log  ( 2 \,  \tr \rho_2)   \big] = \log 2  \qquad 
   \ee
where we used  for $k = 1,2 $
\be
   \tr \rho_k^2 \rho_\av^{-1} = 2 \, \tr \rho_k^{1/2} \Big( \rho_k^{1/2} \frac{1}{\rho_1 + \rho_2}  \rho_k^{1/2}\Big)  \rho_k^{1/2}
       \leq 2 \,  \tr  \rho_k^{1/2} I  \rho_k^{1/2} = 2 \, \tr \rho_k 
\ee
and the inequalities are strict if $\rho_1 \neq \rho_2$ and they are not multiples of pairwise orthogonal projections.
Then we can conclude that for  $x = \half$,  
   \be
      h(\half)  = \log 2  & \geq &  \half \Big[H^\ren_a (\rho_2 , \rho_\av)  +    H^\ren_a(\rho_1 , \rho_\av) \Big] \nn \\
         & \geq & \half \Big[\wtd{H}^\ren_a (\rho_1 , \rho_\av)  +    \wtd{H}^\ren_a(\rho_2 , \rho_\av) \Big]  \\  \nn
         & \geq  &   S(   \rho_\av) - x S(\rho_1) - (1-x) S(\rho_2)
   \ee 
   for all $a \in (1,2] $.   This led us to conjecture that a similar result could be proved for any fixed $x$.
  Milan Mosonyi \cite{M2} then proved the following slightly stronger result and kindly 
   allowed us to include his argument here.
\begin{thm}  {\rm (Mosonyi)}   \label{thm:milan}
   For all $a > 1$,
 \be   \label{newbd}
     h(x)  & \geq  & x \Big[\wtd{H}^\ren_{ a} (\rho_1 , \rho_\av)  + (1-x)   \wtd{H}^\ren_{ a}(\rho_2 , \rho_\av) \Big]  \nn \\   
         & \geq  &   S(   \rho_\av) - x S(\rho_1) - (1-x) S(\rho_2)
 \ee
 \end{thm}  
\noindent{\bf Proof:}  Note that the sandwiched Renyi divergences are monotone increasing in the parameter a, and their limit when 
$ a \raw \infty $ is the max-relative entropy 
$$ H_{\max}(\rho, \gamma):= \inf\{\log\omega: \rho \le \omega \gamma\}.$$
Obviously, 
$\rho_1\le (1/x)\rho_{\av}$, and hence
$$\wtd{H}_a(\rho_1,  \rho_{\av}) \le H_{\max}(\rho_1,  \rho_{\av})\le -\log x  \quad \forall ~ a\in [1/2,+\infty]. $$
Applying the same result to $ \rho_2$,  shows that  \eqref{newbd} holds for every $ a > 1 $.
   
  \bs
  
 \noindent{\bf Acknowledgment:}                
  It is a pleasure to thank Koenraad Audenaert, Fumio Hiai  and Milan Mosonyi for helpful correspondence on an earlier version of the manuscript.  
   IK's research at Perimeter Institute is supported by the Government of Canada through Industry Canada and by the
Province of Ontario through the Ministry of Economic Development and Innovation.     MBR's work was partially supported by NSF Grant CFF-1018401  which is   administered by Tufts University.    MBR would also like to
express her appreciation for the stimulated and hospitable atmosphere at the  Institute for Quantum Computing in Waterloo, where most of this work was done.
\appendix
\section{Numerical Examples}
Since all of the numerical examples were found for qubit density matrices, it is convenient to represent them
using the Bloch sphere representation $\rho=\half [I + \bw \dtsig \big] \equiv  \half [I + \sum_k w_k \sigma_k\big]$
\and identify $\bw_k $ with $\rho_k$. 
\bsq \begin{align}
 \!   \bw_1 & = (0.2876, 0.4322, 0.3112)     & \bw_2  & = (-0.1552, -0.0532, -0.0874)   & x & = 0.7086   \\
   \!    \bw_1 & = (-0.2136, 0.0702, -0.0944)     & \bw_2  & = (-0.5204,0.7790, -0.1772)   & x & = 0.2197\\
 \!     \bw_1 & = (-0.1850,0.7506,-0.6388)     & \bw_2  & = (0.0254,0.0012, 0.0114)   & x & = 0.5218   
\end{align}  \esq
For Example (a),  \eqref{lowbd1} yields a better bound than \eqref{lowbd2}.  
For Example (b), \eqref{lowbd2} yields a better bound  \eqref{lowbd1}.
 It is interesting that even \eqref{lowbd0}, which is stronger than \eqref{lowbd1}, is sometimes weaker than \eqref{lowbd2}. 
 This is the case for Example (c) for which  \eqref{lowbd2} is strictly greater than \eqref{lowbd0}.
 
 {~~}


\begin{thebibliography}{~~}
 
  \bibitem{A} K.M.R. Audenaert 
  ``Quantum Skew Divergence: Theory and Applications''
   	arXiv:1304.5935
 
 \bibitem{B}     S. Beigi  
    ``Sandwiched RŽnyi Divergence Satisfies Data Processing Inequality''
  {\em J. Math. Phys. }{ \bf 54}, 122202 (2013.)   arXiv:1306.5920
  
  \bibitem{Bur}  D. Bures, 
  ``An extension of KakutaniÕs theorem on infinite product measures to the tensor
product of semi-finite w?-algebras''
  {\em  Trans. Amer. Math. Soc. }  {\bf?135} 199--212 (1969).
 
 \bibitem{CL} E. A. Carlen and E.H. Lieb
    ``Remainder Terms for Some Quantum Entropy Inequalities''
      arXiv:1402.3840 
      
     \bibitem{DM}  M. Delbr\"{u}rk and G. Mole\`{i}ere 
     ÔÔStatistische quantenmechanik und thermodynamik,ÕÕ
     {\em Abh. Preuss. Akad.Wissenschaften}
 {\bf 1}, 1--42 (1936).
      

\bibitem{FG}  C. A. Fuchs and J. van de Graaf,
``Cryptographic distinguishability measures for
quantum mechanical states''  {\em  IEEE Trans. Inf. Theory}  {\bf  45} 1216--1227  (1999).

\bibitem{H} F. Hiai, 
``Matrix Analysis and quantum information'' 
S?gaku 65(2) 133--159 (2013)  (Japanese),;
English translation {\em Sugaku Expositions} in press (2014)
 
      
       \bibitem{kim}  I. H. Kim   
      ``Convexity estimate of operator convex functions''
     arXiv:1310.0746 
     
     
 \bibitem{LanR}  O. E. Lanford III and D. W. Robinson,
 ``Mean Entropy of States in Quantum Statistical Mechanics''
   {\em J. Math. Phys. }{ \bf 94},  1120--1125 (1968.) 

     

\bibitem{M} M. Mosonyi 
``Renyi divergences and the classical capacity of finite compound channels''
 arXiv:1310.7525
 
\bibitem{M2} M. Mosonyi, private communication.
        
  \bibitem{MDSFT}        M. M\"{u}ller-Lennert, F. Dupuis, O. Szehr, S. Fehr, and  M. Tomamichel
     ``On quantum Renyi entropies: a new generalization and some properties''
   {\em J. Math. Phys. }{ \bf 54}, 12220e2 (2013.)     arXiv:1306.3142 
  
\bibitem{OP} M. Ohya and D. Petz, {\it Quantum Entropy and Its Use},
(Springer-Verlag, Heidelberg, 1993; 2nd edition 2004).
 
 \bibitem{P} M.S. Pinsker {\it Information and  Information Stability of Random Variables and Processes}
 (Holden Day, 1964).
 
 \bibitem{RFZ}  W. Roga, M. Fannes and K. Zyczkowski
 ``Universal bounds for the Holevo quantity, coherent information and the Jensen-Shannon divergence ''
 `` Phys. Rev. Lett. 105, 040505 (2010).   arXiv:1101.4105
 
 \bibitem{U} A. Uhlmann, 
 ``Density operators an an arena for differential geometry''
 {\em  Rep. Math. Phys.}  {\bf 33} 253--263 (1993).
 
 \bibitem{WWY}  M. M. Wilde, A. Winter, D. Yang 
   ``Strong converse for the classical capacity of entanglement-breaking and Hadamard channels''
     arXiv:1306.1586  
        
 \end{thebibliography}
   \end{document}